\documentclass[conference]{IEEEtran}
\IEEEoverridecommandlockouts
\usepackage{cite}
\usepackage{amsmath,amssymb,amsfonts}
\usepackage{algorithmic}
\usepackage{graphicx}
\usepackage{textcomp}
\usepackage[table]{xcolor}
\usepackage{url}

\usepackage{comment}
\usepackage{booktabs}
\usepackage{multirow}
\usepackage{pifont}

\usepackage[prependcaption,textsize=scriptsize]{todonotes}
\definecolor{mycolor}{HTML}{008000}%
\definecolor{indiagreen}{HTML}{138808}%
\definecolor{papaya}{HTML}{EE892F}%
\definecolor{mygreen}{HTML}{008000}%
\definecolor{myblue}{HTML}{5D8AA8}

\definecolor{mypurplelight}{HTML}{CBC3E3}
\definecolor{mypurple}{HTML}{9966CC}%

\newcommand{\ata}{A$\to$A}
\newcommand{\tta}{T$\to$A}

\newcommand{\yy}{\ding{51}}
\def\BibTeX{{\rm B\kern-.05em{\sc i\kern-.025em b}\kern-.08em
    T\kern-.1667em\lower.7ex\hbox{E}\kern-.125emX}}
\begin{document}

\title{Echoes: A semantically-aligned music deepfake detection dataset}

\author{
\IEEEauthorblockN{Octavian Pascu\IEEEauthorrefmark{1}, Dan Oneata\IEEEauthorrefmark{1}, Horia Cucu\IEEEauthorrefmark{1}, Nicolas M. Müller\IEEEauthorrefmark{2}\IEEEauthorrefmark{3}}
\IEEEauthorblockA{
\IEEEauthorrefmark{1}Politehnica Bucharest, Romania \quad
\IEEEauthorrefmark{2}Fraunhofer AISEC, Germany \quad
\IEEEauthorrefmark{3}Resemble AI\\
\{octavian.pascu,dan\_theodor.oneata,horia.cucu\}@upb.ro, nicolas.mueller@aisec.fraunhofer.de
}
}

\maketitle

\begin{abstract}
    We introduce \textit{Echoes}, a new dataset for music deepfake detection designed for training and benchmarking detectors under realistic and provider-diverse conditions.
    Echoes comprises 4\,468 AI-generated tracks (131 hours of audio) paired with 300 bona-fide references, spanning multiple genres (pop, rock, electronic), and includes content generated by 12 popular AI music generation systems.
    To prevent shortcut learning and promote robust generalization, the dataset is deliberately constructed to be challenging, enforcing semantic-level alignment between spoofed audio and bona-fide references.
    This alignment is achieved by conditioning generated audio samples directly on bona-fide waveforms or song descriptors.

    We evaluate Echoes in a cross-dataset setting against three existing AI-generated music datasets using state-of-the-art Wav2Vec2 XLS-R 2B representations.
    Results show that
    \textit{(i)} Echoes is the hardest in-domain dataset;
    \textit{(ii)} detectors trained on existing datasets transfer poorly to Echoes;
    \textit{(iii)} training on Echoes yields the strongest generalization performance.
    We further introduce complementary protocols on Echoes which show that generalizing to held-out generators remains difficult and that provider identity is more entangled in Echoes than in existing datasets.
    These findings suggest that provider diversity and semantic alignment help learn more transferable detection cues.
    Echoes is released publicly at \url{https://huggingface.co/datasets/Octavian97/Echoes} under \texttt{CC-BY-SA}.
\end{abstract}

\begin{IEEEkeywords}
music deepfake detection, anti-spoofing, AI-generated music, dataset, cross-dataset generalization
\end{IEEEkeywords}

\section{Introduction}
\label{sec:introduction}

Recent advances in music generation have enabled the synthesis of convincing songs that emulate genres, production aesthetics, and vocal styles at scale.
Modern generators can produce realistic music from short prompts, increasing the feasibility of large-scale synthetic music creation.
This shift creates integrity and provenance challenges for the music ecosystem:
synthetic uploads can be used for impersonation, attribution fraud, and automated monetization schemes, while stressing platform-level moderation.
Streaming platforms have reported rapidly growing volumes of fully AI-generated submissions, motivating the need for scalable and reliable detection pipelines \cite{Deezer25-AI28}.

\begin{figure}[t]
  \centering
  \includegraphics[width=0.95\linewidth]{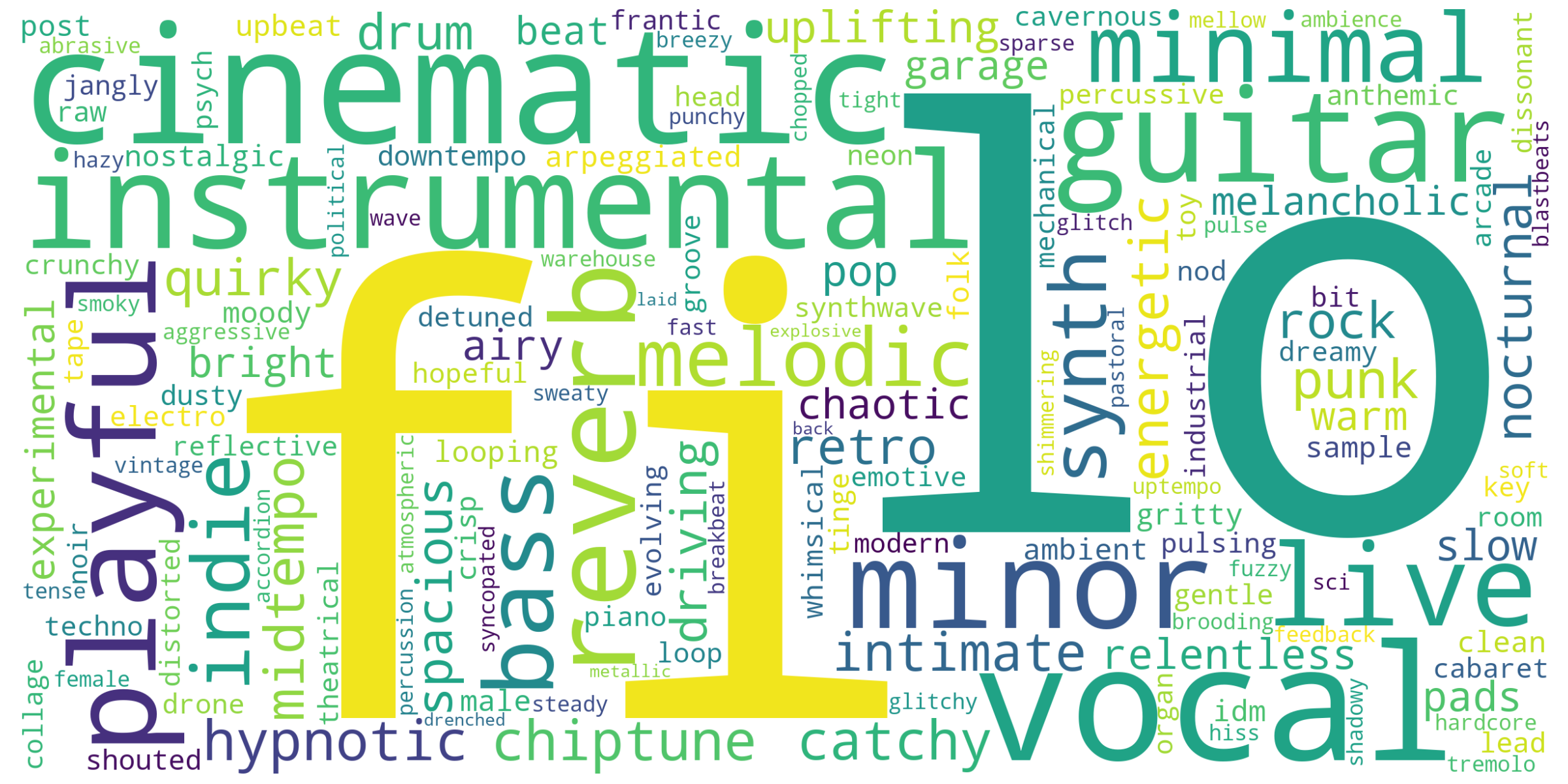}
  \caption{
    Word cloud of generated song descriptions.
    The descriptions are produced by a large language model from songs' metadata and are used to condition music generation in our dataset, Echoes.
  }
  \label{fig:wordcloud}
\end{figure}

Music deepfake detection solutions are emerging in response.
Current approaches are based on techniques from speech anti-spoofing.
These include waveform-based models (e.g., RawNet2) \cite{tak2021end}, spectro-temporal attention models (e.g., AASIST) \cite{jung2022aasist}, and self-supervised encoders (e.g., wav2vec2-XLS-R) paired with lightweight classifiers and data augmentation to improve transfer \cite{chen2024singing, tak2022automatic, pascu2024towards}.
However, as in speech anti-spoofing, generalization often suffers:
models perform well when training and testing data follow similar distributions,
but performance drops sharply when evaluated on unseen generators or different preprocessing pipelines \cite{afchar2024detecting, li2024audio, todisco2019asvspoof, muller2022does}.

An important factor in achieving good generalization is data quality.
Prior work found that datasets can exhibit unwarranted asymmetries \cite{muller2021speech,borzi2022synthetic,smeu2025circumventing},
such as leading silences;
these lead models to learn shortcut features that do not generalize.
One solution is to generate fake samples that are as closely aligned to real samples as possible.
In speech deepfake detection, for example, samples have been passed through vocoders \cite{wang2023spoofed};
while in image deepfake detection, real images have been reconstructed using a latent diffusion model \cite{rajan2025}.
This deliberate alignment encourages detectors to focus on the generation artifacts rather than spurious differences.

Data diversity is equally important.
Different generator architectures and generation techniques produce distinct artifacts.
This makes detectors trained on one generator fail on another.
Current architectures vary significantly, from autoregressive modeling over audio tokens \cite{agostinelli2023musiclm, copet2023simple}
to latent diffusion models \cite{liu2023audioldm, evans2025stable}.
These methods produce differences in timbre, mixing, and post-processing, further complicating generalization \cite{cros2025ai, li2024audio}.
In addition, the generation process itself introduces challenges \cite{zang2024singfake, zhang2024svdd, afchar2024detecting}:
a detector should flag a single forged component (e.g., synthesized vocals over real accompaniment),
as well as an entirely generated song.

However, existing deepfake music datasets do not emphasize these aspects,
and as such there remains a gap between existing benchmarks and realistic evaluation requirements.
Among the existing options,
AIME \cite{Groetschla25-AIME} targets large-scale human preference studies on short excerpts, prioritizing subjective evaluation over detector training;
FakeMusicCaps \cite{comanducci2024fakemusiccaps} supports detection and attribution but remains largely clip-oriented and does not contain long-form structure;
SONICS \cite{rahman2024sonics} emphasizes long-duration counterfeit-song detection, yet covers only two popular providers.

To address these gaps, we introduce \textit{Echoes}, an evaluation-centric dataset for training and benchmarking AI-generated music detectors.
Echoes is designed to be semantically aligned and covers a broad range of providers.
The dataset contains $4\,468$ AI-generated tracks totaling 131 hours and spanning pop, rock, and electronic genres, built from 300 bona-fide songs sourced from the Free Music Archive \cite{defferrard2016fma}.
To preserve high-level semantics, %
for each bona-fide track we generate song-specific descriptors with an LLM (Figure~\ref{fig:wordcloud}) and use these descriptions (and, where supported, reference audio) to prompt 12 popular music generators.
Table~\ref{tab:ai_music_datasets} summarizes the existing datasets under three axes (provider diversity, scale, and duration) and shows that Echoes fills in an important benchmarking gap.

Finally, we report initial baseline results on our dataset using a detector built on self-supervised Wav2Vec2 features.
In cross-dataset experiments, we observe the strongest transfer when training on Echoes and testing on the three datasets.
This result highlights the value of a more diverse and better aligned training dataset.
Going beyond the cross-dataset setting, we also evaluate detection of held-out (unseen) generators through a leave-one-provider-out protocol, and quantify how separable the providers are through an attribution task.

To summarize, our main contributions are:
\textit{(i)} a new semantically-aligned deepfake music dataset which contains both short-form and long-form synthetic songs;
\textit{(ii)} coverage across major generator families by sourcing deepfakes from 12 providers;
\textit{(iii)} a cross-dataset baseline, reported with uncertainty estimates, showing substantial generalization gaps and improved transfer when training on Echoes; and
\textit{(iv)} complementary evaluation protocols on Echoes: unseen-provider detection, a conditioning-modality analysis, and provider attribution.

\section{Dataset description}
\label{sec:dmdd}

\begin{table}[t]
  \caption{Comparison of AI music datasets in terms of provider diversity, scale, and song duration.}
  \label{tab:ai_music_datasets}
  \centering
  \setlength{\tabcolsep}{4pt}
  \renewcommand{\arraystretch}{1.08}
  \newcommand{\mypm}[1]{\footnotesize $\pm$#1}
  \begin{tabular}{lrrr@{\hskip 0.01in}l}
    \toprule
    & & \multicolumn{3}{c}{\bf Duration} \\
    \cmidrule(lr){3-5}
    \textbf{Name} & \bf Providers & \textbf{Total (h)} & \multicolumn{2}{r}{\bf Average (s)} \\
    \midrule
    AIME~\cite{Groetschla25-AIME}                    & 7  & 58   &  10.18 & \mypm{0.3}  \\
    SONICS~\cite{rahman2024sonics}                   & 2  & 4\,751  & 144.00 & \mypm{57.0} \\
    FakeMusicCaps~\cite{comanducci2024fakemusiccaps} & 5  & 77    &  10.08 & \mypm{0.1}  \\
    Echoes (ours)                                    & 12 & 131   & 105.36 & \mypm{74.1} \\
    \bottomrule
  \end{tabular}
\end{table}

\subsection{Generation pipeline}

To build our dataset,
we start from bona-fide music tracks from the Free Music Archive (FMA) \cite{defferrard2016fma}.
We select 300 tracks licensed under CC0, CC-BY or the public domain.
For each bona-fide track, we generate corresponding fake samples using models from 12 providers (Table~\ref{tab:ai_music_providers}).

To obtain aligned samples that preserve the original style, we use each track's title and genre to prompt \textit{GPT-5 Thinking}
to extract a list of stylistic characteristics.
We query the large language model with the following prompt:

\begin{quote}\small\it
Describe the following track's characteristics in as few words as possible with the idea of using the descriptions to generate a deepfake audio as close sounding as possible. Use commas between characteristics.
\end{quote}

\begin{figure}[t]
  \centering
  \includegraphics[width=0.95\linewidth]{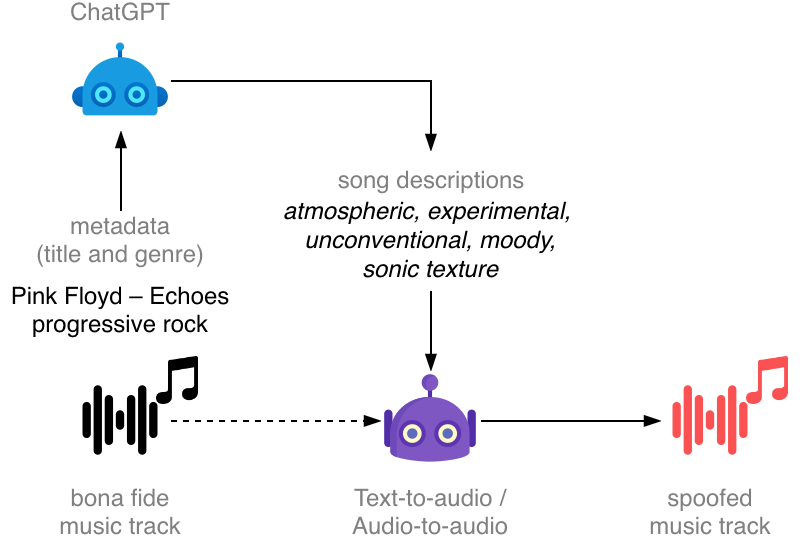}
  \caption{%
    Generation pipeline for the Echoes dataset.
    Spoofed music tracks are generated from corresponding bona-fide audio in two ways:
    (1) by directly passing the bona-fide track through an audio-to-audio model;
    (2) by providing its metadata (title and genre) to an LLM to produce stylistic descriptions, which is then used to condition a text-to-audio model.
  }
  \label{fig:music_gen}
\end{figure}

The resulting descriptions are then used to condition the \textit{text-to-audio} generation process.
For providers that support reference-audio conditioning, we also generate an \emph{audio-to-audio} sample using the same descriptors together with the original track as reference.
This pipeline is illustrated in Figure~\ref{fig:music_gen} and yields 4\,468 generated (fake) tracks.
Each provider contributes one text-to-audio sample per bona-fide reference, and the five providers that support reference-audio conditioning (AudioLDM, DiffRhythm, MusicGen, Riffusion, SongGen) additionally contribute audio-to-audio samples.
The resulting per-provider totals (149 to 594 tracks; Table~\ref{tab:ai_music_providers}) deviate from these nominal targets due to generation failures, provider-side duration limits, and usage constraints, which explains why the dataset size is not a simple multiple of the number of bona-fide tracks.

\begin{table*}[t]
  \caption{
    AI-music providers used to create Echoes. ``Unknown'' means the provider does not publicly specify the model.
    The two generator types are either audio-to-audio (\ata) or text-to-audio (\tta).
  }
  \label{tab:ai_music_providers}
  \centering
  \renewcommand{\arraystretch}{1.08}
  \newcommand{\mypm}[1]{\color{gray} \tiny $\pm$#1}
  \begin{tabular}{l l c c c r r@{\hskip 0.01in}l r}
    \toprule
    &
    &
    &
    \multicolumn{2}{c}{\bf Generator} &
    \bf Number &
    \multicolumn{2}{r}{\bf Average} &
    \bf Total \\
    \bf Provider &
    \bf Model &
    \bf Commercial &
    \bf \tta &
    \bf \ata &
    \bf songs &
    \multicolumn{2}{r}{\bf duration (s)} &
    \bf duration (h) \\
    \midrule
    AceStep~\cite{gong2025ace}                    & \tt ACE-Step-v1-3.5B    & No  & \yy &     & 294 &  66.27 & \mypm{20.53} &  5.4 \\
    AudioLDM~\cite{liu2023audioldm}               & \tt audioldm-s-full     & No  & \yy & \yy & 587 &  30.72 & \mypm{0.00}  &  5.0 \\
    Brev~\cite{Brev26}                            & \tt v4.5 Pro            & Yes & \yy &     & 298 & 190.67 & \mypm{62.44} & 15.8 \\
    DiffRhythm~\cite{ning2025diffrhythm}          & \tt DiffRhythm-v1.2     & Yes & \yy & \yy & 594 & 121.16 & \mypm{14.60} & 20.0 \\
    ElevenLabs~\cite{ElevenLabs26}                & \tt v2                  & Yes & \yy &     & 300 & 188.98 & \mypm{30.61} & 15.7 \\
    Mubert~\cite{Mubert26}                        & Unknown                 & Yes & \yy &     & 149 & 220.27 & \mypm{28.28} &  9.1 \\
    MusicGen~\cite{copet2023simple}               & \tt musicgen-melody     & No  & \yy & \yy & 591 &  30.00 & \mypm{0.00}  &  4.9 \\
    Riffusion~\cite{Forsgren22-Riffusion}         & \tt FUZZ 2.0            & Yes & \yy & \yy & 300 & 178.97 & \mypm{39.49} & 14.9 \\
    SongGen~\cite{liu2025songgen}                 & \tt SongGen\_mixed\_pro & Yes & \yy & \yy & 561 &  28.13 & \mypm{3.13}  &  4.4 \\
    Stable Audio~\cite{StabilityAI26-StableAudio} & \tt AudioSparx 2.0      & Yes & \yy &     & 194 & 180.00 & \mypm{0.00}  &  9.7 \\
    Suno~\cite{Suno26}                            & \tt v5                  & Yes & \yy &     & 300 & 179.30 & \mypm{54.84} & 14.9 \\
    Udio~\cite{Udio26}                            & Unknown                 & Yes & \yy &     & 300 & 130.94 & \mypm{0.12}  & 10.9 \\
    \bottomrule
  \end{tabular}
\end{table*}

\subsection{Additional metadata}

Apart from the audio files and their binary labels, we provide additional metadata to support a broader range of research tasks.
Specifically, for each sample we include the generator name, generator type (text-to-audio or audio-to-audio), musical genre, and the prompt used for generation.
This metadata facilitates reproducibility and enables richer analyses, such as
attributing samples to the corresponding provider,
studying the impact of the conditioning modality, or
developing genre-aware models.

\subsection{Dataset statistics}

Echoes contains 4\,468 generated tracks totaling approximately 131 hours, alongside the 300 bona-fide reference tracks.
Compared to existing datasets (Table~\ref{tab:ai_music_datasets}), our dataset is larger than AIME (58 hours) and FakeMusicCaps (77 hours).
Only SONICS contains more data (4\,751 hours), but it is less diverse as it includes samples from just two providers.

\begin{figure}[t]
  \centering
  \includegraphics[width=\linewidth]{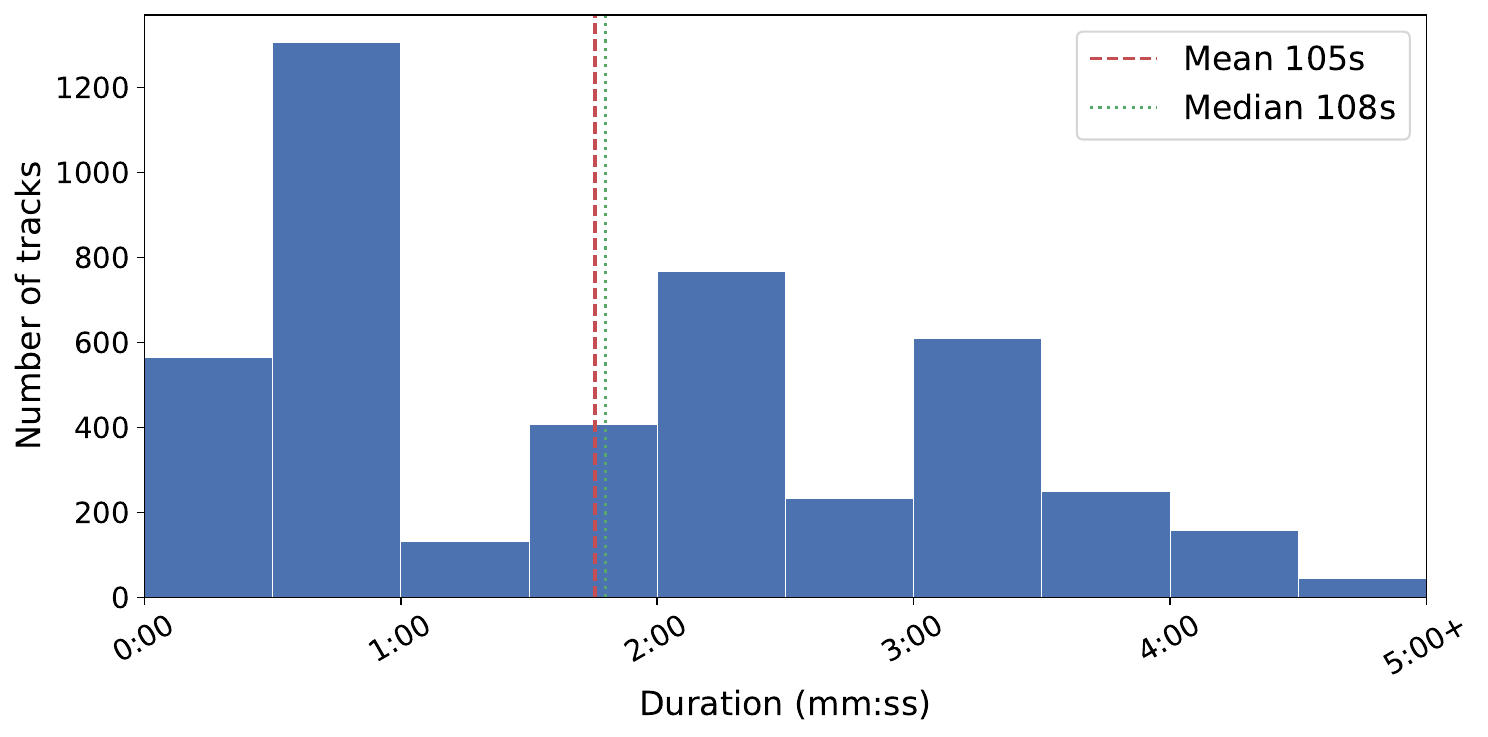}
  \caption{Track duration distribution in Echoes.
  }
  \label{fig:dur_distrib}
\end{figure}

\textbf{Duration distribution.}
Track durations extend up to 480 seconds, with a median length of 108 seconds (Figure~\ref{fig:dur_distrib}).
The spikes in the distribution appear from providers that
generate fixed-length outputs (around 30\,s):
for example, AudioLDM generates audio clips of 30.72\,s on average, MusicGen outputs clips of exactly 30\,s, while SongGen outputs clips of 28.13\,s
(Table~\ref{tab:ai_music_providers}).

\textbf{Per-provider statistics.}
Table~\ref{tab:ai_music_providers} lists the 12 providers used to generate the samples, along with the underlying model (when available), number of generated tracks, average duration, and commercial status.
Overall, nine out of the 12 systems are commercial, and the number of tracks per provider ranges from 149 to 594.

\begin{table*}[!ht]
    \caption{
    Equal error rate (EER; \%, lower is better) when training and testing on different combinations of datasets, reported as mean $\pm$ standard deviation over five random track-level splits.
    Gray diagonal entries are in-domain results; green cells mark the best out-of-domain result per column (training set) and the best out-of-domain average.
    }
    \label{tab:res}
    \centering
    \setlength{\tabcolsep}{5pt}
    \renewcommand{\arraystretch}{1.15}
    \newcommand{\mypm}[1]{\color{gray} \tiny $\pm$#1}
    \begin{tabular}{l rrrr c}
        \toprule
        \multirow{2}{*}{\textbf{Training datasets $\downarrow$}} &
        \multicolumn{4}{c}{\textbf{Test dataset $\rightarrow$}} &
        \multirow{2}{*}{\textbf{Average (Out-of-domain)}} \\
        \cmidrule(lr){2-5}
        & AIME & SONICS & FakeMusicCaps & Echoes & \\
        \midrule
        AIME &
        \textcolor{black!45}{9.85 \mypm{0.61}} &
        18.71 \mypm{2.19} &
        \cellcolor{green!15}\textbf{19.98 \mypm{0.39}} &
        \cellcolor{green!15}\textbf{30.68 \mypm{2.25}} &
        23.12 \\
        SONICS &
        24.85 \mypm{0.96} &
        \textcolor{black!45}{4.79 \mypm{1.08}} &
        36.02 \mypm{0.96} &
        31.03 \mypm{2.23} &
        30.63 \\
        FakeMusicCaps &
        30.04 \mypm{1.17} &
        38.66 \mypm{2.19} &
        \textcolor{black!45}{9.18 \mypm{0.96}} &
        42.34 \mypm{1.81} &
        37.01 \\
        Echoes (ours) &
        \cellcolor{green!15}\textbf{23.07 \mypm{1.00}} &
        \cellcolor{green!15}\textbf{15.24 \mypm{1.48}} &
        25.12 \mypm{1.16} &
        \textcolor{black!45}{\textbf{15.32 \mypm{0.72}}} &
         \cellcolor{green!15}\textbf{21.14} \\
        \bottomrule
    \end{tabular}
\end{table*}

\section{Evaluation}

We provide initial results on our dataset, as well as cross-dataset generalization results using the three other major available datasets (AIME, SONICS, FakeMusicCaps).
We complement the cross-dataset analysis with three analyses defined on Echoes: unseen-provider detection (leave-one-provider-out), a per-provider and conditioning-modality study, and provider attribution.
Unless stated otherwise, all results are reported as mean $\pm$ standard deviation over five random train--test splits.

\subsection{Training setup}

\hspace*{\parindent}\textbf{Data processing.}
All audio files are split into non-overlapping
10-second segments.
This design choice is motivated by several reasons.
First, AIME and FakeMusicCaps are clip-based and use 10-second excerpts;
segmenting all corpora to 10-second clips ensures a fair and standardized protocol across datasets.
Second, 10-second segments provide a simple baseline that measures how much discriminative information is available from short local cues.
Importantly, Echoes still differs from clip-only datasets because these segments are sampled from full-length songs and span diverse providers and generation settings, which increases variability even under a fixed-window evaluation.
We view long-context modeling as a complementary direction and leave sequence-level evaluation on full tracks for future work.

\textbf{Train-test split.}
We split the data into 80:20 training and evaluation sets at the \emph{track level} for all datasets: all 10-second segments extracted from the same full-length track are kept in the same split, which prevents train--test leakage.
We repeat each experiment over five random splits and report mean $\pm$ standard deviation.

\textbf{Detection model.}
As the detection baseline, we use a self-supervised learning (SSL) front-end coupled with a lightweight classifier.
We extract 1\,920-dimensional embeddings using Wav2Vec2 XLS-R 2B \cite{baevski2020wav2vec,babu2021xls}, which has shown strong performance in audio deepfake detection and related singing spoofing settings \cite{pascu2024towards,chen2024singing}.
We keep the Wav2Vec2 XLS-R 2B encoder \emph{frozen} and obtain a fixed-dimensional embedding for each 10-second segment by applying a temporal average pooling over the frame-level representations from the final layer.

On top of these embeddings, we train a logistic regression classifier to predict whether each segment is bona-fide or AI-generated by minimizing binary cross-entropy with L2 regularization.
We use \texttt{scikit-learn} logistic regression with a maximum of 5\,000 iterations and $C=10^6$ (all other parameters are set to default), resulting in 1\,921 trainable parameters (1\,920 weights plus a bias).
Performance is measured in terms of equal error rate (EER), a threshold-free metric commonly used in spoofing detection.

\subsection{Cross-dataset evaluation}

The model is trained on each dataset in turn and evaluated on all four datasets (Table~\ref{tab:res}).
This setup measures \textit{(i)} how easy each dataset is in-domain (train and test on the same dataset) and \textit{(ii)} how well a model trained on one dataset transfers to the others.

\subsubsection{In-domain detection}
We first look at the diagonal of Table~\ref{tab:res}, which represents in-domain results.
We observe that Echoes is the most challenging dataset under the same model and training protocol:
the classifier achieves an EER of 15.32\%, which exceeds those obtained on AIME (9.85\%), FakeMusicCaps (9.18\%), and SONICS (4.79\%).
From a dataset perspective, this is a desirable property: higher in-domain EER suggests fewer exploitable shortcuts and a more demanding benchmark for developing robust detection methods.

\subsubsection{Out-of-domain detection}
We next examine the off-diagonal entries in Table~\ref{tab:res}, which represent out-of-domain testing.
The first observation is that the out-of-domain performance is much lower than in-domain. This behavior is consistent with the broader audio deepfake detection literature, where strong in-domain performance does not correlate with out-of-domain performance \cite{afchar2024detecting, li2024audio, todisco2019asvspoof, muller2022does}.
Models trained on AIME, SONICS, or FakeMusicCaps perform poorly on Echoes, with EERs of 30.7\%, 31.0\%, and 42.3\%, respectively.
This indicates that cues learned from clip-centric datasets or low-diversity setups do not carry over well to Echoes.
Notably, the poor transfer from AIME cannot be attributed to unseen generators alone: six of AIME's seven provider families also appear in Echoes (albeit partly with different model versions; see Section~\ref{sec:attribution}), yet the AIME-trained model still reaches 30.7\% EER.
This suggests that the difficulty of Echoes stems from its semantic alignment and long-form construction rather than from provider novelty alone.

In the other direction, training on Echoes yields the best average out-of-domain performance (21.1\% average EER, compared to 23.1\% for the next best training set), and it achieves the best transfer to AIME (23.1\%) and SONICS (15.2\%) compared to the other training choices.
Overall, these results support the goal of Echoes: it is difficult in-domain and it encourages learning cues that generalize better than those learned from the other benchmarks.

\subsection{Unseen-provider evaluation}
\label{sec:lopo}

The cross-dataset comparison above conflates several factors at once (genre coverage, preprocessing, and provider sets).
To isolate generalization to \emph{unseen generators} we run a leave-one-provider-out (LOPO) protocol within Echoes.
For each provider, we train the detector on the bona-fide training tracks together with the fake samples of the remaining 11 providers, and evaluate on the held-out bona-fide tracks together with the fake samples of the excluded provider.
Bona-fide tracks are split 80:20 at the track level, and results are averaged over five random bona-fide splits.

To contextualize the held-out results, we also compute a \emph{seen} per-provider EER: the standard in-domain detector (trained on all 12 providers) is evaluated separately on each provider's test fakes against the bona-fide test tracks.
Averaged over providers, the seen EER is 15.06\%, consistent with the overall in-domain result of 15.32\% (the small difference stems from weighting providers equally rather than by segment count).

Table~\ref{tab:lopo} reports the seen and unseen EERs side by side.
Detecting an unseen provider is markedly harder: the mean unseen EER is 22.33\%, a gap of $+7.3$ percentage points over the seen setting.
More interestingly, the cost of being unseen varies widely across generators: DiffRhythm, Suno, and Brev lose only 2--3 percentage points when held out, whereas Mubert, Stable Audio, AudioLDM, and MusicGen degrade by 10--11 percentage points, indicating that detection artifacts transfer well within some generator families but poorly across others.
The hardest providers are largely consistent in both regimes (Mubert and Stable Audio rank last either way), while DiffRhythm remains the easiest even when unseen.
These results show that, despite the broad provider coverage of Echoes, unseen-generator detection is far from solved, and we propose LOPO on Echoes as a standardized protocol for measuring progress on this axis.

\begin{table}[t]
  \caption{Per-provider detection EER (\%) on Echoes when the provider is \emph{seen} during training versus \emph{unseen} (leave-one-provider-out).
  Seen: the in-domain detector is evaluated on each provider's test fakes against the bona-fide test tracks.
  Unseen: the detector is trained on the remaining 11 providers.
  Mean $\pm$ standard deviation over five splits; $\Delta$ is the unseen$-$seen gap.}
  \label{tab:lopo}
  \centering
  \setlength{\tabcolsep}{4pt}
  \renewcommand{\arraystretch}{1.08}
  \newcommand{\mypm}[1]{\color{gray} \tiny $\pm$#1}
  \begin{tabular}{l r@{\hskip 0.01in}l r@{\hskip 0.01in}l r}
    \toprule
    \textbf{Provider} & \multicolumn{2}{c}{\textbf{Seen}} & \multicolumn{2}{c}{\textbf{Unseen}} & \multicolumn{1}{c}{$\boldsymbol{\Delta}$} \\
    \midrule
    DiffRhythm   &  8.95 & \mypm{1.00} & 11.38 & \mypm{1.23} &  $+2.43$ \\
    MusicGen     & 12.62 & \mypm{1.16} & 22.81 & \mypm{1.62} & $+10.19$ \\
    SongGen      & 12.67 & \mypm{0.85} & 20.63 & \mypm{1.88} &  $+7.96$ \\
    AceStep      & 13.34 & \mypm{0.81} & 19.44 & \mypm{2.41} &  $+6.10$ \\
    Suno         & 13.45 & \mypm{0.91} & 16.45 & \mypm{1.65} &  $+3.00$ \\
    Riffusion    & 13.89 & \mypm{1.09} & 18.57 & \mypm{1.52} &  $+4.68$ \\
    Brev         & 13.92 & \mypm{1.11} & 17.05 & \mypm{2.22} &  $+3.13$ \\
    Udio         & 15.56 & \mypm{0.87} & 24.88 & \mypm{1.57} &  $+9.32$ \\
    ElevenLabs   & 16.69 & \mypm{0.46} & 25.08 & \mypm{1.37} &  $+8.39$ \\
    AudioLDM     & 17.39 & \mypm{1.35} & 27.48 & \mypm{1.72} & $+10.09$ \\
    Stable Audio & 20.50 & \mypm{1.69} & 31.09 & \mypm{1.37} & $+10.59$ \\
    Mubert       & 21.69 & \mypm{1.01} & 33.16 & \mypm{1.40} & $+11.47$ \\
    \midrule
    \textbf{Mean} & \textbf{15.06} & & \textbf{22.33} & & $\boldsymbol{+7.27}$ \\
    \bottomrule
  \end{tabular}
\end{table}

\subsection{Effect of the conditioning modality}
\label{sec:tta_ata}

Echoes contains generations conditioned in two ways: on text descriptors only (\tta) or additionally on the bona-fide reference audio (\ata).
Since five providers support both modes (Table~\ref{tab:ai_music_providers}), the dataset enables a controlled comparison of how the conditioning modality affects detection difficulty.
We evaluate the in-domain detector separately on the \tta{} and \ata{} test fakes of each dual-mode provider, against the bona-fide test tracks and report the results in Table \ref{tab:tta_ata}.
For all five dual-mode providers, \ata{} generations are harder to detect than their \tta{} counterparts, with a mean EER increase of $+2.06\%$ (14.15\% vs 12.10\%).
This is consistent with the design intent of the dataset: conditioning on the bona-fide waveform produces generations that inherit more of the reference's acoustic characteristics, reducing the separability between bona-fide and generated content.

\begin{table}[t]
  \caption{Detection EER (\%) per conditioning modality, matched within provider, for the five providers supporting both text-to-audio (\tta) and audio-to-audio (\ata) generation.}
  \label{tab:tta_ata}
  \centering
  \setlength{\tabcolsep}{6pt}
  \renewcommand{\arraystretch}{1.08}
  \begin{tabular}{l r r r}
    \toprule
    \textbf{Provider} & \textbf{\tta} & \textbf{\ata} & \multicolumn{1}{c}{$\boldsymbol{\Delta}$} \\
    \midrule
    AudioLDM   & 16.61 & 17.84 & $+1.23$ \\
    DiffRhythm &  7.80 & 10.18 & $+2.38$ \\
    MusicGen   & 12.47 & 12.86 & $+0.39$ \\
    Riffusion  & 11.76 & 16.15 & $+4.39$ \\
    SongGen    & 11.84 & 13.75 & $+1.91$ \\
    \midrule
    \textbf{Mean} & \textbf{12.10} & \textbf{14.15} & $\boldsymbol{+2.06}$ \\
    \bottomrule
  \end{tabular}
\end{table}

\subsection{Provider attribution}
\label{sec:attribution}

The metadata released with Echoes also enables \emph{provider attribution}: identifying which generator produced a given fake sample.
Beyond being a useful forensic task in itself, attribution quantifies how separable the providers are in the representation space.
We train a multiclass logistic regression (same hyper-parameters and frozen SSL embeddings as before) on the fake segments only, predicting the generating provider, using a single track-level split stratified per provider.
For comparison, we apply the same protocol to AIME (seven providers), FakeMusicCaps (five providers), and SONICS (two providers).
Attribution is defined at the provider (generator-family) level: where a dataset's metadata exposes finer model versions (e.g., Suno v3/v3.5 or Stable Audio v1/v2 in AIME, and the chirp/udio variants in SONICS), we collapse them into a single provider class so that label sets are comparable across datasets.

Table~\ref{tab:attribution} summarizes the results.
On Echoes, the classifier reaches 78.6\% accuracy over 12 classes, with per-class recall as low as 0.60 for Brev and 0.64 for Suno.
In contrast, attribution is close to ceiling on the other datasets: 94.8\% on AIME (seven classes), 91.0\% on FakeMusicCaps (five classes), and 99.0\% on SONICS (two classes).
Notably, a larger label set alone does not explain the gap: AIME separates its seven providers almost perfectly, whereas Echoes shows substantial confusion between providers despite using the same features and classifier.
Provider identity is therefore largely recoverable from the SSL features, but the providers in Echoes are considerably more entangled than in existing datasets, which is consistent with its broader provider coverage and aligned generation conditions.

\begin{table}[t]
  \caption{Provider attribution with a multiclass logistic regression on frozen SSL embeddings (fake segments only, track-level split). Chance is the majority-class baseline (accuracy of always predicting the most frequent provider).}
  \label{tab:attribution}
  \centering
  \setlength{\tabcolsep}{4pt}
  \renewcommand{\arraystretch}{1.08}
  \begin{tabular}{l r r r r}
    \toprule
    \textbf{Dataset} & \textbf{Classes} & \textbf{Chance (\%)} & \textbf{Accuracy (\%)} & \textbf{Macro-F1} \\
    \midrule
    Echoes (ours) & 12 & 15.1 & 78.6 & 0.82 \\
    AIME          &  7 & 73.0 & 94.8 & 0.89 \\
    FakeMusicCaps &  5 & 20.0 & 91.0 & 0.91 \\
    SONICS        &  2 & 61.6 & 99.0 & 0.99 \\
    \bottomrule
  \end{tabular}
\end{table}

\begin{figure}[t]
  \centering
  \includegraphics[width=\linewidth]{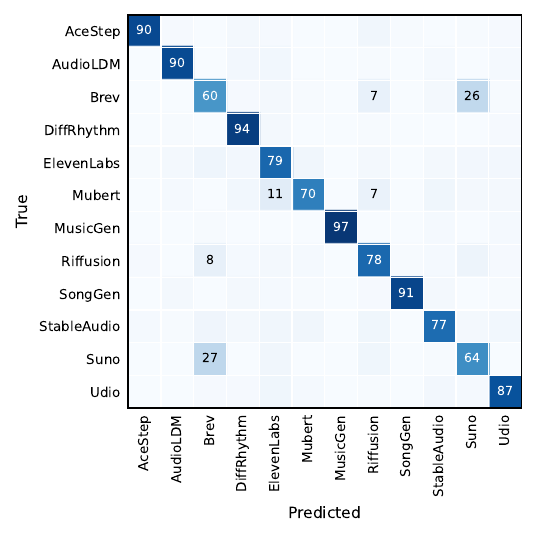}
  \caption{Row-normalized confusion matrix (\%) of the 12-way provider attribution on Echoes (rows: true provider; columns: predicted). Cells below 5\% are left unannotated.}
  \label{fig:confusion}
\end{figure}

Figure~\ref{fig:confusion} shows the row-normalized confusion matrix on Echoes.
DiffRhythm is the most separable class (0.94 recall), consistent with it also being the easiest provider to detect both seen and unseen (Table~\ref{tab:lopo}).
At the other end, Brev (0.60), Suno (0.64), and Mubert (0.69) are frequently mistaken for other providers.
The dominant confusion is mutual between Brev and Suno: 26\% of Brev segments are attributed to Suno and 27\% of Suno segments to Brev, more than twice of any other provider pair, suggesting closely related generation pipelines or production characteristics. This also offers an explanation to why Brev and Suno did not see a performance drop in the LOPO evaluation: training on one tool offers good cues to detect generated music with the other tool. The next largest confusions involve Mubert being mistaken for ElevenLabs (11\%) and the Brev--Riffusion pair (7--8\% in each direction).

\section{Conclusion}
\label{sec:conclusion}

We presented \textit{Echoes}, a semantically-aligned dataset for AI-generated music detection built to support training and benchmarking under realistic provider diversity.
Echoes contains $4\,468$ AI-generated tracks ($131$ hours) paired with $300$ bona-fide tracks across pop, rock, and electronic music genres.
For each bona-fide song from the Free Music Archive, we generate AI-produced counterparts from 12 popular music generators using LLM-derived, song-specific descriptors and, when available, audio-conditioned generation. This pairing strategy aims to keep high-level intent (style/theme) similar between bona-fide and AI-generated tracks, making the AI-generated music detection task harder and reducing simple content-based shortcuts. In summary, \textit{Echoes} is the most diverse AI-generated music dataset to date.

Our experiments show three key results.
First, Echoes is the hardest dataset in-domain under the same SSL+linear baseline ($15.3\%$ EER), compared with AIME ($9.9\%$), SONICS ($4.8\%$), and FakeMusicCaps ($9.2\%$).
Second, cross-dataset transfer is weak in general, and especially weak when testing on Echoes: models trained on AIME, SONICS, or FakeMusicCaps reach 30.7--42.3\% EER on Echoes. In contrast, training on Echoes yields the strongest average out-of-domain performance ($21.1\%$ EER), suggesting that combining semantic alignment with multi-provider coverage pushes detectors toward cues that transfer better across datasets.
Third, our additional protocols on Echoes show that detecting held-out generators remains difficult (mean leave-one-provider-out EER of $22.3\%$) and that provider identity is more entangled in Echoes (78.6\% attribution accuracy over 12 classes) than in existing datasets (91--99\%).
We further find that audio-conditioned (\ata) generations are consistently harder to detect than their text-conditioned counterparts ($+2.1\%$ EER on matched providers), providing detection-based support for the alignment strategy.

Echoes is released under \texttt{CC-BY-SA} to enable more realistic evaluation of AI-generated music detection and to encourage reporting beyond in-domain results.
All bona-fide material originates from permissively licensed sources (CC0, CC-BY, or public domain), and the dataset is intended solely for research on the detection, attribution, and benchmarking of AI-generated music.
Future work will evaluate detectors under settings closer to deployment (e.g., common audio post-processing and partial or mixed real/AI content).

\bibliographystyle{IEEEtran}
\bibliography{mybib}

@inproceedings{chen2024singing,
  title={Singing Voice Graph Modeling for SingFake Detection},
  author={Chen, Xuanjun and Wu, Haibin and Jang, Roger and Lee, Hung-yi},
  booktitle={Proc. Interspeech 2024},
  pages={4843--4847},
  year={2024}
}

@inproceedings{defferrard2016fma,
  title={FMA: A Dataset For Music Analysis},
  author={Defferrard, Micha{\"e}l and Benzi, Kirell and Vandergheynst, Pierre and Bresson, Xavier},
  booktitle={18th International Society for Music Information Retrieval Conference},
  year={2017}
}

@inproceedings{liu2023audioldm,
  title={AudioLDM: Text-to-Audio Generation with Latent Diffusion Models},
  author={Liu, Haohe and Chen, Zehua and Yuan, Yi and Mei, Xinhao and Liu, Xubo and Mandic, Danilo and Wang, Wenwu and Plumbley, Mark},
  booktitle={Proceedings of the 40th International Conference on Machine Learning, PMLR 2023},
  volume={202},
  pages={21450--21474},
  year={2023},
  organization={International Machine Learning Society (IMLS)}
}

@article{copet2023simple,
  title={Simple and controllable music generation},
  author={Copet, Jade and Kreuk, Felix and Gat, Itai and Remez, Tal and Kant, David and Synnaeve, Gabriel and Adi, Yossi and D{\'e}fossez, Alexandre},
  journal={Advances in Neural Information Processing Systems},
  volume={36},
  pages={47704--47720},
  year={2023}
}

@inproceedings{Groetschla25-AIME,
  title={Benchmarking Music Generation Models and Metrics via Human Preference Studies},
  author={Gr{\"o}tschla, Florian and Solak, Ahmet and Lanzend{\"o}rfer, Luca A and Wattenhofer, Roger},
  booktitle={ICASSP},
  year={2025},
}

@inproceedings{rahman2024sonics,
  title={SONICS: Synthetic Or Not-Identifying Counterfeit Songs},
  author={Rahman, Md Awsafur and Hakim, Zaber Ibn Abdul and Sarker, Najibul Haque and Paul, Bishmoy and Fattah, Shaikh Anowarul},
  booktitle={The Thirteenth International Conference on Learning Representations},
  year={2024}
}

@misc{Deezer25-AI28,
    author       = {{Deezer}},
    title        = {{Deezer: 28\% of all delivered music is now fully AI-generated}},
    howpublished = {{Deezer Newsroom (press release)}},
    month        = sep,
    year         = 2025,
    note         = {Accessed: 2026-01-11},
    url          = {https://newsroom-deezer.com/2025/09/28-fully-ai-generated-music/}
}

@article{afchar2024detecting,
  title={Detecting music deepfakes is easy but actually hard},
  author={Afchar, Darius and Meseguer-Brocal, Gabriel and Hennequin, Romain},
  journal={arXiv preprint arXiv:2405.04181},
  year={2024}
}

@article{cros2025ai,
  title={The {AI} Music Arms Race: On the Detection of {AI}-Generated Music},
  author={Cros Vila, Laura and Sturm, Bob and Casini, Luca and Dalmazzo, David},
  journal={Transactions of the International Society for Music Information Retrieval},
  volume={8},
  number={1},
  pages={179--194},
  year={2025}
}

@article{li2024audio,
  title={From Audio Deepfake Detection to {AI}-Generated Music Detection--{A} Pathway and Overview},
  author={Li, Yupei and Milling, Manuel and Specia, Lucia and Schuller, Bj{\"o}rn W},
  journal={arXiv preprint arXiv:2412.00571},
  year={2024}
}

@inproceedings{zang2024singfake,
  title={Singfake: Singing voice deepfake detection},
  author={Zang, Yongyi and Zhang, You and Heydari, Mojtaba and Duan, Zhiyao},
  booktitle={ICASSP},
  year={2024},
}

@inproceedings{zhang2024svdd,
  title={{SVDD} 2024: The inaugural singing voice deepfake detection challenge},
  author={Zhang, You and Zang, Yongyi and Shi, Jiatong and Yamamoto, Ryuichi and Toda, Tomoki and Duan, Zhiyao},
  booktitle={SLT},
  year={2024},
}

@article{agostinelli2023musiclm,
  title={Music{LM}: Generating music from text},
  author={Agostinelli, Andrea and Denk, Timo I and Borsos, Zal{\'a}n and Engel, Jesse and Verzetti, Mauro and Caillon, Antoine and Huang, Qingqing and Jansen, Aren and Roberts, Adam and Tagliasacchi, Marco and others},
  journal={arXiv preprint arXiv:2301.11325},
  year={2023}
}

@inproceedings{evans2025stable,
  title={Stable audio open},
  author={Evans, Zach and Parker, Julian D and Carr, CJ and Zukowski, Zack and Taylor, Josiah and Pons, Jordi},
  booktitle={ICASSP},
  year={2025},
}

@article{comanducci2024fakemusiccaps,
    author       = {Luca Comanducci and Paolo Bestagini and Stefano Tubaro},
    journal      = {Journal of Imaging},
    number       = 7,
    pages        = {242},
    title        = {{FakeMusicCaps}: A Dataset for Detection and Attribution of Synthetic Music Generated via Text-to-Music Models},
    volume       = 11,
    year         = 2025,
    doi          = {10.3390/jimaging11070242},
}

@inproceedings{todisco2019asvspoof,
  title={ASVspoof 2019: Future Horizons in Spoofed and Fake Audio Detection},
  author={Todisco, Massimiliano and Wang, Xin and Vestman, Ville and Sahidullah, Md and Delgado, Hector and Nautsch, Andreas and Yamagishi, Junichi and Evans, Nicholas and Kinnunen, Tomi and Lee, Kong Aik},
  booktitle={Interspeech 2019},
  pages={1008--1012},
  year={2019},
  organization={International Speech Communication Association}
}

@inproceedings{muller2022does,
  title={Does audio deepfake detection generalize?},
  author={M{\"u}ller, Nicolas M and Czempin, Pavel and Dieckmann, Franziska and Froghyar, Adam and B{\"o}ttinger, Konstantin},
  booktitle={Interspeech},
  year={2022}
}

@inproceedings{tak2021end,
  title={End-to-end anti-spoofing with {RawNet2}},
  author={Tak, Hemlata and Patino, Jose and Todisco, Massimiliano and Nautsch, Andreas and Evans, Nicholas and Larcher, Anthony},
  booktitle={ICASSP},
  year={2021},
}

@inproceedings{tak2022automatic,
  title={Automatic Speaker Verification Spoofing and Deepfake Detection Using Wav2vec 2.0 and Data Augmentation},
  author={Tak, Hemlata and Todisco, Massimiliano and Wang, Xin and Jung, Jee-weon and Yamagishi, Junichi and Evans, Nicholas},
  booktitle={The Speaker and Language Recognition Workshop (Odyssey 2022)},
  year={2022},
  organization={ISCA}
}

@inproceedings{jung2022aasist,
  title={{AASIST}: Audio anti-spoofing using integrated spectro-temporal graph attention networks},
  author={Jung, Jee-weon and Heo, Hee-Soo and Tak, Hemlata and Shim, Hye-jin and Chung, Joon Son and Lee, Bong-Jin and Yu, Ha-Jin and Evans, Nicholas},
  booktitle={ICASSP},
  year={2022},
}

@inproceedings{pascu2024towards,
  title={Towards generalisable and calibrated audio deepfake detection with self-supervised representations},
  author={Pascu, Octavian and Stan, Adriana and Oneata, Dan and Oneata, Elisabeta and Cucu, Horia},
  booktitle={Interspeech},
  volume={2024},
  pages={4828--4832},
  year={2024}
}

@misc{Forsgren22-Riffusion,
    author       = {{Seth Forsgren and Hayk Martiros}},
    title        = {{Riffusion -- Stable Diffusion for Real-Time Music Generation}},
    howpublished = {{GitHub repository}},
    url          = {https://github.com/riffusion/riffusion-hobby},
    month        = dec,
    year         = 2022,
    note         = {{Accessed: 2026-01-13}},
}

@misc{StabilityAI26-StableAudio,
    author       = {{Stability AI}},
    title        = {{Stable Audio 2.5}},
    howpublished = {{Website}},
    url          = {https://stability.ai/stable-audio},
    month        = jan,
    year         = 2026,
    note         = {{Accessed: 2026-01-13}},
}

@misc{Mubert26,
    author       = {{Mubert}},
    title        = {{Mubert {AI} Music Generator}},
    howpublished = {{Website}},
    url          = {https://mubert.com/},
    month        = jan,
    year         = 2026,
    note         = {{Accessed: 2026-01-13}},
}

@misc{Brev26,
    author       = {{Brev}},
    title        = {{Brev {AI} Music Generator}},
    howpublished = {{Website}},
    url          = {https://brev.ai/},
    month        = jan,
    year         = 2026,
    note         = {{Accessed: 2026-01-13}},
}

@misc{ElevenLabs26,
    author       = {{ElevenLabs}},
    title        = {{ElevenLabs {AI} Music Generator}},
    howpublished = {{Website}},
    url          = {https://ElevenLabs.io/},
    month        = May,
    year         = 2026,
    note         = {{Accessed: 2026-05-13}},
}

@misc{Udio26,
    author       = {{Udio}},
    title        = {{Udio: {AI} Music Generator}},
    howpublished = {{Website}},
    url          = {https://www.udio.com/},
    month        = jan,
    year         = 2026,
    note         = {{Accessed: 2026-01-13}},
}

@article{gong2025ace,
  title={{ACE}-step: A step towards music generation foundation model},
  author={Gong, Junmin and Zhao, Sean and Wang, Sen and Xu, Shengyuan and Guo, Joe},
  journal={arXiv preprint arXiv:2506.00045},
  year={2025}
}

@article{ning2025diffrhythm,
  title={Diff{R}hythm: Blazingly fast and embarrassingly simple end-to-end full-length song generation with latent diffusion},
  author={Ning, Ziqian and Chen, Huakang and Jiang, Yuepeng and Hao, Chunbo and Ma, Guobin and Wang, Shuai and Yao, Jixun and Xie, Lei},
  journal={arXiv preprint arXiv:2503.01183},
  year={2025}
}

@inproceedings{liu2025songgen,
  title={SongGen: A Single Stage Auto-regressive Transformer for Text-to-Song Generation},
  author={Liu, Zihan and Ding, Shuangrui and Zhang, Zhixiong and Dong, Xiaoyi and Zhang, Pan and Zang, Yuhang and Cao, Yuhang and Lin, Dahua and Wang, Jiaqi},
  booktitle={International Conference on Machine Learning},
  pages={38351--38364},
  year={2025},
  organization={PMLR}
}

@misc{Suno26,
    author       = {{Suno}},
    title        = {{Suno: {AI} Music}},
    howpublished = {{Website}},
    url          = {https://suno.com/},
    month        = jan,
    year         = 2026,
    note         = {{Accessed: 2026-01-13}},
}

@inproceedings{babu2021xls,
  title={XLS-R: Self-supervised Cross-lingual Speech Representation Learning at Scale},
  author={Babu, Arun and Wang, Changhan and Tjandra, Andros and Lakhotia, Kushal and Xu, Qiantong and Goyal, Naman and Singh, Kritika and von Platen, Patrick and Saraf, Yatharth and Pino, Juan and others},
  booktitle={Proc. Interspeech 2022},
  pages={2278--2282},
  year={2022}
}

@inproceedings{baevski2020wav2vec,
  title={wav2vec 2.0: A framework for self-supervised learning of speech representations},
  author={Baevski, Alexei and Zhou, Yuhao and Mohamed, Abdelrahman and Auli, Michael},
  booktitle={NeurIPS},
  year={2020}
}

@inproceedings{rajan2025,
  title={On the Effectiveness of Dataset Alignment for Fake Image Detection},
  author={Rajan, Anirudh Sundara and Ojha, Utkarsh and Schloesser, Jedidiah and Lee, Yong Jae},
  year={2025},
  booktitle={ICLR},
}

@inproceedings{borzi2022synthetic,
  title={Is synthetic voice detection research going into the right direction?},
  author={Borz{\`\i}, Stefano and Giudice, Oliver and Stanco, Filippo and Allegra, Dario},
  booktitle={CVPRW},
  year={2022}
}

@inproceedings{muller2021speech,
  title={Speech is Silver, Silence is Golden: What do ASVspoof-trained Models Really Learn?},
  author={M{\"u}ller, Nicolas and Dieckmann, Franziska and Czempin, Pavel and Canals, Roman and B{\"o}ttinger, Konstantin and Williams, Jennifer},
  booktitle={Proc. ASVSPOOF 2021},
  pages={55--60},
  year={2021}
}

@inproceedings{smeu2025circumventing,
  title={Circumventing shortcuts in audio-visual deepfake detection datasets with unsupervised learning},
  author={Smeu, Stefan and Boldisor, Dragos-Alexandru and Oneata, Dan and Oneata, Elisabeta},
  booktitle={CVPR},
  year={2025}
}

@inproceedings{wang2023spoofed,
  title={Spoofed training data for speech spoofing countermeasure can be efficiently created using neural vocoders},
  author={Wang, Xin and Yamagishi, Junichi},
  booktitle={ICASSP},
  year={2023},
}

\end{document}